\documentclass[pra, twocolumn, floatfix]{revtex4}
\usepackage{graphicx}
\usepackage{color}
\usepackage{amsmath, amsfonts, amssymb, bm}
\begin{document}

\title{Two-color phase-of-the-phase spectroscopy applied to nonperturbative\\ 
electron-positron pair production in strong oscillating electric fields}
\author{J. Bra{\ss},$^1$ R. Milbradt,$^1$ S. Villalba-Ch\'avez,$^1$ G. G. Paulus,$^2$ and C. M\"uller$^1$}
\address{$^1$Institut f\"ur Theoretische Physik I, Heinrich-Heine-Universit\"at D\"usseldorf, Universit\"atsstr. 1, 40225 D\"usseldorf, Germany\\
$^2$Institute of Optics and Quantum Electronics, Friedrich Schiller University, Max-Wien-Platz 1, 07743 Jena, Germany}
\date{\today}
\begin{abstract}
Production of electron-positron pairs from vacuum in strong bichromatic electric fields, oscillating in time with a fundamental frequency and its second harmonic, is studied. Strong-field processes occuring in such field configurations are generally known to be sensitive to the relative phase between the field modes.
Phase-of-the-phase spectroscopy has recently been introduced in the context of strong-field photoionization  as a systematic means to analyze these coherence effects. We apply this method to field-induced pair production by calculating the phase dependence of the momentum-resolved particle yields. We show that asymmetric checkerboard patterns arise in the phase-of-the-phase spectra, similarly to those found in strong-field photoionization. The physical origin of these characteristic structures, which differ between the created electron and positron, are discussed.
\end{abstract}

\maketitle

\section{Introduction}
In the presence of very strong electric fields, the quantum vacuum can become unstable and decay into electron-positron ($e^-e^+$) pairs. This was first predicted for constant electric fields \cite{Sauter, Schwinger} and later extended to electric fields harmonically oscillating in time \cite{Brezin, Popov1, Popov2}. Pair production can also be induced by high-intensity laser fields. While the vacuum remains stable in the presence of a plane electromagnetic wave due its vanishing field invariants \cite{Schwinger}, the combination of a laser wave with, for example, a $\gamma$-ray photon, a charged particle or another counterpropagating laser wave may lead to pair production \cite{Reviews}. The problem of field-induced pair production has found renewed interest in recent years because it comes into experimental reach at upcoming high-intensity laser facilities, such as the Extreme-Light Infrastructure \cite{ELI}, the Exawatt Center for Extreme Light Studies \cite{XCELS} or the European X-Ray Free-Electron Laser \cite{XFEL}. 

Various interaction regimes of pair production in a monochromatic oscillating field can be distinguished by the value of the dimensionless parameter $\xi = |e|E_0/(mc\omega)$, with electric field amplitude $E_0$, oscillation frequency $\omega$, electron charge $e$ and mass $m$, and speed of light $c$ \cite{Reviews}. 
For $\xi\ll 1$, the pair production rate obeys a perturbative multiphoton power law. For $\xi\gg 1$ (providing $E_0$ stays subcritical), it displays instead a nonperturbative exponential field dependence, resembling the case of constant electric field \cite{Schwinger}. In between these asymptotic domains lies the nonperturbative regime of intermediate coupling strengths $\xi\sim 1$, where analytical treatments of the problem are rather difficult. Noteworthy, a close analogy with strong-field photoionization in intense laser fields exists where the corresponding regimes of perturbative multiphoton ionization, tunneling ionization and above-threshold ionization (ATI) are well known.

Very pronounced effects may be triggered when an oscillating field contains two frequencies $\omega_1$ and $\omega_2$. Strong amplification of pair production has been predicted to occur in fields consisting of a strong low-frequency and a weak high-frequency component, with $\omega_1\ll\omega_2\sim mc^2/\hbar$ \cite{Schutzhold, DiPiazza, Orthaber, JansenPRA, Grobe2, Akal, Kampfer}. Another interesting situation arises when the two frequencies are commensurate, for example $\omega_2=2\omega_1$. Then characteristic quantum interferences between different multiphoton pathways occur, along with a dependence on the relative phase between the field modes. This was demonstrated for pair production in the superposition of a high-energy gamma-photon and an intense laser wave \cite{Narozhny, Jansen} as well as in combined laser and nuclear Coulomb fields \cite{Krajewska, Augustin, Roshchupkin}. The relative phase was shown here to exhibit a distinct influence on the momentum distribution of created particles. Similar coherence effects were found for pair production in trains of electric field pulses due to multiple-slit interferences in the time domain \cite{Dunne2, slit, combs, modulation, grating, Granz}. 

Two-color quantum interferences and relative-phase effects are well established in intense laser interactions with atoms and molecules \cite{Ehlotzky}. Forming the basis for coherent phase control, they allow to specifically manipulate strong-field processes. This way it is, for example,  possible to influence photoelectron yields from strong-field ionization \cite{SFPI}, enhance the efficiency of high-harmonic generation \cite{HHG}, and spatially direct the fragmentation of molecules in photodissociation \cite{dissociation}.

Recently, phase-of-the-phase spectroscopy has been developed as novel method to analyze relative-phase effects in two-color strong-field phenomena \cite{BauerPRL, BauerJPB, BauerPRA, Wuerzler}. It relies on the observation that, in mathematical terms, a relative phase is a continuous variable and a bichromatic field is a $2\pi$-periodic function thereof. This periodic property is passed on to observables, which are derived from the field, such as momentum distributions of particles. The observable of interest can therefore be expanded into Fourier series, with the relative-phase dependence being encoded in the complex Fourier coefficients. The latter are given by their absolute value and complex phase (which is thus ``the phase of the phase''). The method was applied to the tunneling \cite{BauerPRL} and above-threshold \cite{BauerJPB} regimes of strong-field photoionization with linearly polarized fields in joint experimental and theoretical studies. It has also been extended to two-color fields of circular \cite{BauerPRA} or mutually orthogonal polarization \cite{Wuerzler}.

In the present paper, we study $e^-e^+$ pair production by bifrequent oscillating electric fields in the nonperturbative regime with $\xi\sim 1$. The corresponding time-dependent Dirac equation is solved numerically to obtain the production probabilities for given particle momenta. Our focus lies on the relative-phase dependence of the momentum distributions which is analyzed by phase-of-the-phase spectroscopy. The resulting spectra are shown to exhibit a characteristic checkerboard pattern, similar to those obtained in corresponding studies of strong-field photoionization \cite{BauerPRL, BauerJPB, BauerPRA}. The method offers a possibility to distinguish in future high-intensity laser experiments coherent pair production channels from incoherent background processes, which might arise from residual rest gas atoms, for example.

Our paper is organized as follows. In Sec.~II we briefly outline our theoretical approach to the problem which was derived in detail previously. Our numerical results regarding the relative-phase dependence of bichromatic field-induced pair production are presented in Sec.~III. Concluding remarks are given in Sec.~IV. Relativistic units with $\hbar=c=4\pi\epsilon_0=1$ are used throughout unless otherwise stated.

\section{Theoretical framework}
Our goal is to analyze the relative-phase dependence of pair production in an oscillating electric field comprising a fundamental frequency $\omega$ and its second harmonic $2\omega$. The field is chosen to be linearly polarized in $y$-direction. In temporal gauge, such a field $\vec E(t)=-\dot{\vec A}(t)$ can be described by a vector potential of the form
\begin{eqnarray}
\vec{A}(t) = \big[ A_1\sin(\omega t) +A_2\sin(2\omega t+\varphi) \big]\,F(t)\,\vec{\rm e}_y\ ,
\label{A}
\end{eqnarray}
where $A_j$ is the amplitude of the $j$-th mode ($j\in\{1,2\}$) and $\varphi$ is the relative phase between the modes. Besides, the field is confined by an envelope function 
\begin{eqnarray}
F(t) = \left\{ \begin{array}{ll}
\sin^2\!\left(\frac{1}{2}\omega t\right) &,\ \ 0\le t < \tau \\
1 &,\ \ \tau \le t < T-\tau \\
\sin^2\!\left(\frac{1}{2}\omega t\right) &,\ \ T-\tau \le t\le T\\
0 &,\ \ \mbox{otherwise} \end{array} \right.
\end{eqnarray}
with turn-on and turn-off increments $\tau = \frac{\pi}{\omega}$ of half-cycle duration each and a plateau of constant intensity in between. The plateau region comprises $N$ oscillation cycles of the fundamental mode, so that the total duration of the electric field pulse is $T=(N+1)\frac{2\pi}{\omega}$.

The pair production probability in a time-dependent electric field can be obtained by solving a coupled system of ordinary differential equations \cite{Akal, Kampfer, Mostepanenko, Gitman, Hamlet, Mocken}. We use the following representation which was derived in \cite{Mocken,Hamlet}:
\begin{eqnarray}
\dot{f}(t) &=& \kappa(t)f(t) + \nu(t)g(t)\ ,\nonumber\\
\dot{g}(t) &=& -\nu^*(t)f(t) + \kappa^*(t)g(t)\ ,
\label{system}
\end{eqnarray}
with 
\begin{eqnarray}
\kappa(t) &=& ieA(t)\,\frac{p_y}{\varepsilon_{\vec{p}}}\ ,\nonumber\\
\nu(t) &=& -ieA(t)\,e^{2i\varepsilon_{\vec{p}} t}\,\left[ \frac{(p_x-ip_y)p_y}{\varepsilon_{\vec{p}}(\varepsilon_{\vec{p}}+m)} + i\, \right]\ .
\label{nu}
\end{eqnarray}
It is obtained from the time-dependent Dirac equation when an ansatz of the form $\psi_{\vec p}(\vec r,t) = f(t)\, \phi_{\vec p}^{(+)}(\vec r,t) + g(t)\, \phi_{\vec p}^{(-)}(\vec r,t)$ is inserted. Here, $\phi_{\vec p}^{(\pm)}\sim e^{i(\vec p\cdot\vec r \mp \varepsilon_{\vec{p}} t)}$, with $\varepsilon_{\vec{p}}=\sqrt{{\vec p}^{\,2}+m^2}$, denote free Dirac states with momentum $\vec p$ and positive or negative energy. The suitability of this ansatz relies first of all on the fact that in a spatially homogeneous external field, according to Noether's theorem, the canonical momentum is conserved. Since the latter coincides with the kinetic momentum $\vec p$ of a free particle outside the time intervall when the field is on, it is possible to treat the invariant subspace spanned by the usual four free Dirac states with momentum $\vec p$ separately. Because of the rotational symmetry of the problem about the field axis, the momentum vector can be parametrized as $\vec p = (p_x,p_y,0)$ with transversal (longitudinal) component $p_x$ ($p_y$). As a consequence, one can find a conserved spin-like operator, which allows to reduce the effective dimensionality of the problem further from four to two basis states \cite{Mocken, Hamlet}.

In accordance with the ansatz mentioned above, the time-dependent coefficients $f(t)$ and $g(t)$ describe the occupation amplitudes of a positive-energy and negative-energy state, respectively. The system of differential equations \eqref{system} is solved with the initial conditions $f(0)=0$, $g(0)=1$. After the field has been switched off, $f(T)$ represents the occupation amplitude of an electron state with momentum $\vec p$, positive energy $\varepsilon_{\vec{p}}$ and certain spin projection. Taking the two possible spin degrees of freedom into account, we obtain the probability for creation of a pair with given momentum as 
\begin{eqnarray}
W(\vec p, \varphi) = 2\,|f(T)|^2\ .
\label{W}
\end{eqnarray}
From the manifold parameters which the pair production probability $W$ depends on, our notation highlights the electron momentum $\vec p$ and the relative phase $\varphi$ because they are of main interest here. Note that the created positron has momentum $-\vec p$, so that the total momentum of each pair vanishes.

As explicated in the introduction, the function $W(\vec p, \varphi)$ is $2\pi$-periodic in the phase variable $\varphi$. It can therefore be expanded into Fourier series according to
\begin{eqnarray}
W(\vec p, \varphi) &=& \sum_{\ell = -\infty}^\infty W_\ell(\vec p\,)\ e^{i\ell\varphi}\nonumber \\
&=& W_0(\vec p\,) + 2\sum_{\ell = 1}^\infty |W_\ell(\vec p\,)| \cos[\ell\varphi+\Phi_\ell(\vec p\,)]\ .\nonumber\\
\label{sum}
\end{eqnarray}
The Fourier coefficients can be expressed as $W_\ell = |W_\ell|\,e^{i\Phi_\ell}$, with the absolute value $|W_\ell|$ (called ``relative-phase contrast'' in \cite{BauerPRL,BauerJPB,BauerPRA}) and the complex phase $\Phi_\ell$ (called ``phase of the phase''). These quantities will be investigated in the next section.

Before moving on to the numerical results, we remark that -- under suitable conditions -- strong electric fields oscillating in time can serve as simplified models for intense laser pulses. Particularly in the regime of relativistic intensities which are relevant here \cite{dipole}, the field resulting from the superposition of two counterpropagating laser waves, sharing the same frequency, amplitude and polarization direction, may be represented by an electric background oscillating in time, provided the characteristic pair formation length is much smaller than the laser wavelength and focusing scale. At high frequencies ($\omega\gtrsim 0.1m$), however, significant differences between pair production in an oscillating electric field and pair production in a standing laser wave arise from the spatial dependence and magnetic component of the latter \cite{Ruf, Alkofer2, Dresden, Grobe3}. In our numerical investigations we will apply field frequencies of this magnitude, for reasons of computational feasibility. The corresponding outcomes can therefore be transferred to the case of laser fields only in a qualitative manner. Nevertheless, some general features of the phase-of-the-phase spectra discussed below 
are expected to find their counterparts in laser-induced pair production as well.

\section{Results and Discussion}

\subsection{Choice of Field Parameters}
We have applied the method of phase-of-the-phase spectroscopy to pair production by a strong bifrequent electric field in the nonperturbative multiphoton regime. Previous studies based on monofrequent fields have revealed that the pair production shows characteristic resonances whenever the ratio between the energy gap and the field frequency attains integer values \cite{Popov1, Popov2, Mocken, Bauke, Grobe1}. The energy gap is given by $2\bar{\varepsilon}$, with the time-averaged particle quasi-energies
\begin{eqnarray}
\bar{\varepsilon} = \frac{1}{T}\int_0^T\sqrt{m^2+p_x^2+[p_y - eA(t)]^2}\,dt\ ,
\label{epsilon}
\end{eqnarray}
where $\vec p$ denotes the electron momentum \cite{q_0}. For example, in a monofrequent field with $\xi=1$ one obtains $\bar{\varepsilon}\approx 1.21m$ for vanishing particle momenta ($\vec p=0$). The enhancement as compared with the corresponding field-free energy [$\varepsilon_{\vec{p}}=m$ for $\vec p=0$] is a result of field dressing. As a consequence, a field frequency of $\omega=0.49072m$ leads to resonant production of particles at rest by absorption of five field quanta (``photons'') \cite{Mocken}. To allow for a comparison of our results with this earlier study (see also the recent analysis of pair production in electric double pulses \cite{Granz}), we have used the same frequency value in our numerical calculations. Besides, the normalized amplitude of the fundamental mode is taken as $\xi_1=|e|A_1/m =1$, lying in the nonperturbative multiphoton regime of pair production. The second harmonic mode is chosen with $\xi_2=|e|A_2/m\sim 0.1$. Since the parameter $\xi$ corresponds to the inverse of the Keldysh parameter $\gamma$ for strong-field photoionization, our field parameters are closely related to 
those of Ref.~\cite{BauerJPB} where $\gamma\approx 1$ for the fundamental mode and $A_2/A_1=0.05$.
We point out that the resonant nature of pair production in an oscillating electric field constitutes a difference to the nonresonant process of strong-field photoionization. In the latter case, the electron momentum is not conserved because the atomic nucleus generates a space-dependent field and can absorb recoil momentum.

\begin{figure}[t]  
\begin{center}
\includegraphics[width=0.42\textwidth]{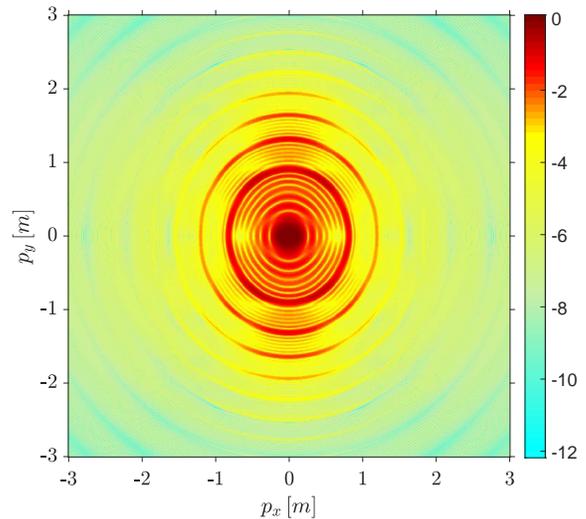}
\end{center}
\vspace{-0.5cm} 
\caption{Two-dimensional momentum distribution of the electron (or positron) created in a bifrequent electric field with $\xi_1=1$, $\xi_2=0.1$, $N=7$, $\omega=0.49072m$ and $\varphi=0$ [see Eq.\,\eqref{A}]. The polarization direction corresponds to the $p_y$ axis. The center around $p_x=p_y=0$ is determined by a $5\omega$ resonance and the characteristic ring structure arises from higher multiphoton resonances with total energy absorption of $6\omega$, $7\omega$, $8\omega$, etc. The color coding refers to $\log_{10}W(\vec p,\varphi)$.}
\label{figure2D}
\end{figure}

The pair production probability \eqref{W} resulting for the chosen field parameters is shown in Fig.~\ref{figure2D}, in dependence on the momenta $p_x$ and $p_y$ of one of the created particles. For the chosen value of the relative phase ($\varphi=0$), the distribution is mirror symmetric under the transformations $p_x\to-p_x$ and $p_y\to-p_y$. This implies, in particular, that the distributions for electrons and positrons coincide in this case. A characteristic ring structure of multiphoton resonances can be seen (similarly to ATI rings in strong-field photoionization). It arises from the fact that, under resonant conditions, the pair production probability \eqref{W} as function of the interaction time $T$ exhibits Rabi-like oscillations between the negative- and positive-energy Dirac continua with maximum amplitude of 2. For the chosen interaction time, which corresponds to $N=7$ cycles of the fundamental mode, this maximum amplitude is reached for the $5\omega$ resonance at the center of Fig.~\ref{figure2D}. 
The resonance condition, in the present case of a bifrequent field, reads $(n_1+2n_2)\omega=2\bar{\varepsilon}$. We note that, in contrast to the monofrequent case \cite{Mocken} and a bifrequent, but noncommensurate situation \cite{Akal}, there are several quantum pathways which contribute to a specific resonance. For example, a total energy of $5\omega$ can be absorbed by either $n_1=5$ low-frequency photons from the fundamental mode, $n_1=3$ low-frequency photons and $n_2=1$ high-frequency photon from the second harmonic mode, or $n_1=1$ low-frequency photon and $n_2=2$ high-frequency photons. These various pathways interfere, which generates a dependence of the pair production probability on the relative phase between the field modes.

Before moving on to the next subsection, we would like to mention that quantum interferences can lead to visible signatures in monochromatic fields as well. For example, distinct carpetlike structures have been observed in the momentum spectra of ATI photoelectrons under certain emission directions \cite{Korneev}. They arise from interfering contributions to the ionization yield, which correspond to two different emission times during a single field cycle and lead to the same electron momentum. A similar substructure of alternating maxima and minima along the resonance rings has also been found in the momentum distributions of electron-positron pairs produced by monofrequent electric fields \cite{Popov2} (for an illustration, see Fig.~11 in \cite{Mocken}). In contrast to these phenomena, phase-of-the-phase spectroscopy allows to study  
{\it changes} in observables, such as momentum distributions, when a controllable phase parameter is externally varied.

\subsection{Phase Dependence of Pair Production}

\begin{figure}[b]  
\vspace{-0.25cm}
\begin{center}
\includegraphics[width=0.45\textwidth]{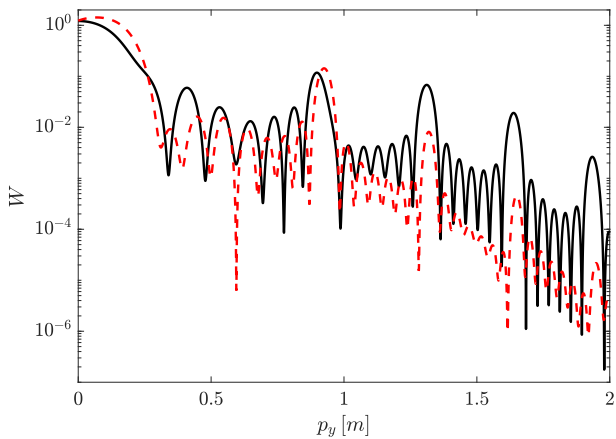}
\end{center}
\vspace{-0.5cm} 
\caption{Longitudinal momentum distribution of electrons created in a bifrequent electric field with $\xi_1=1$, $\xi_2=0.1$, $N=7$ and $\omega=0.49072m$ [see Eq.\,\eqref{A}]. The black solid (red dashed) curve refers to a relative phase of $\varphi=0$ ($\varphi=\frac{\pi}{2}$). The transverse momentum vanishes, $p_x=0$.}
\label{figure1D}
\end{figure}

An illustration of the relative phase dependence of pair production in a bifrequent electric field is depicted in Fig.~\ref{figure1D}. It shows the production probability for two different phase values, when the electron momentum along the field direction varies from 0 to $2m$ while its transverse momentum is kept fixed to zero. Qualitatively, both curves look similar, starting from maximum values of probability and showing a falling tendency, with pronounced multiphoton resonance peaks in between. Quantitatively, however, there are clear differences. For example, compared to the outcome for $\varphi=0$ (black solid curve), more particles are produced with small momenta around $p_y\approx 0.2m$ whereas much less particles have momenta above $\approx 1.2m$ when the phase is chosen as $\varphi=\frac{\pi}{2}$ (red dashed curve).
Similar phase effects arise in the transversal momentum distributions. Furthermore, the positions of the resonance peaks in Fig.~\ref{figure1D} are slightly shifted when the relative phase is varied. This can be understood by noting that the precise form of the vector potential enters into the quasi-energy \eqref{epsilon}. As a consequence, the latter exhibits a weak dependence on $\varphi$, so that the resonance condition is fulfilled at slightly shifted particle momenta. In the range of field parameters considered here, these shifts are on the order of $10^{-2}m$ \cite{Milbradt}. 

\begin{figure}[t]  
\begin{center}
\includegraphics[width=0.45\textwidth]{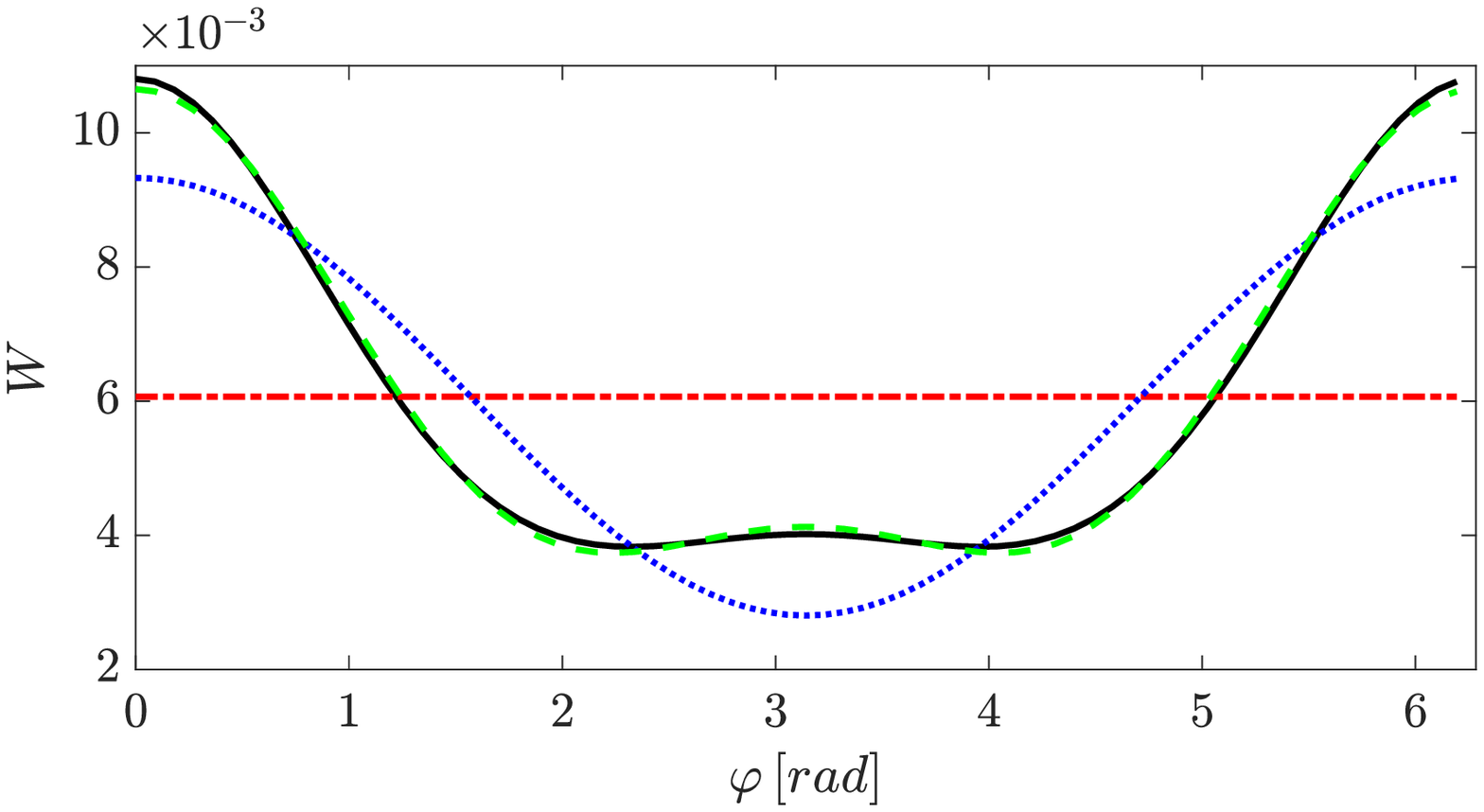}\\
\vspace{1.3cm}
\includegraphics[width=0.45\textwidth]{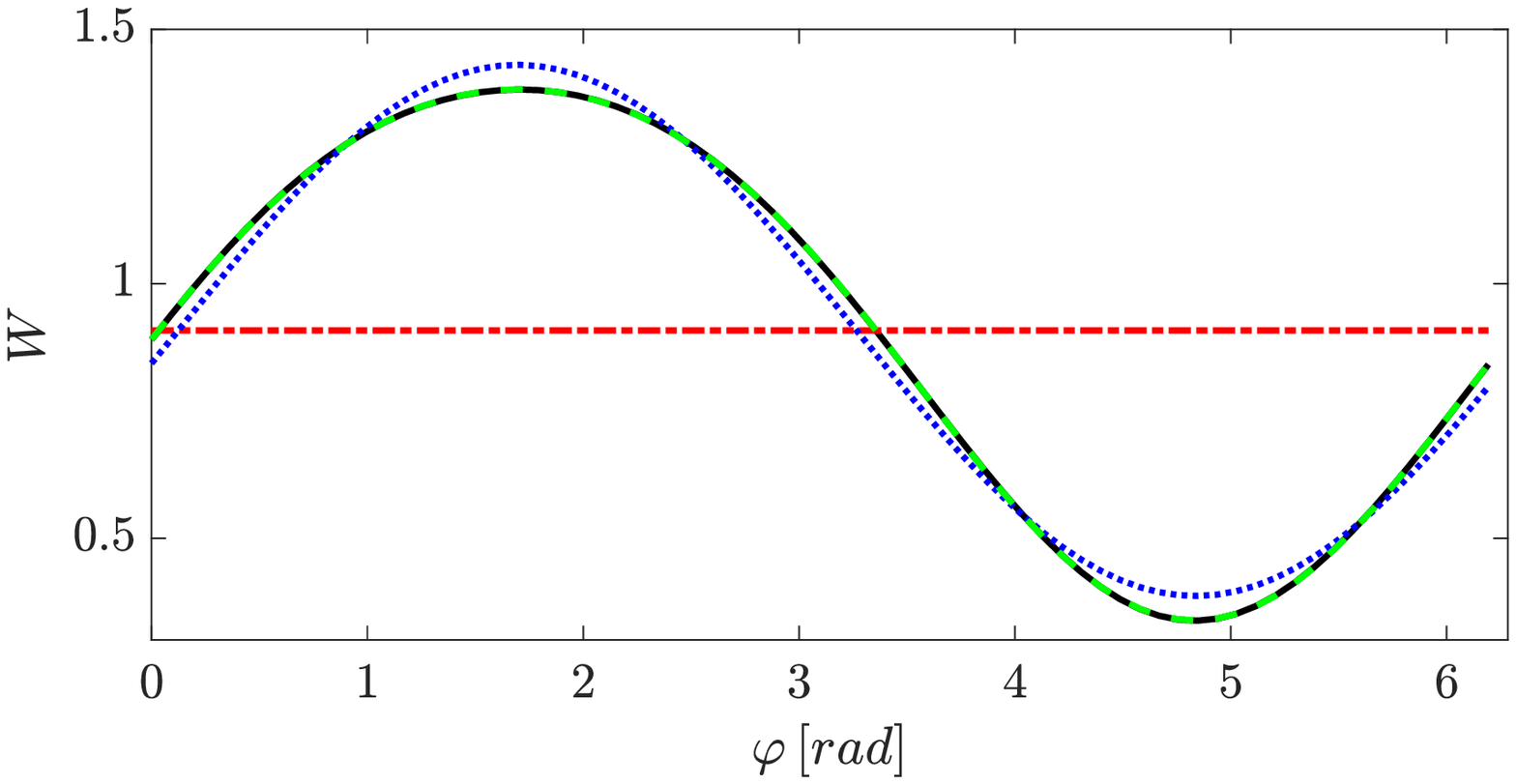}
\end{center}
\vspace{0.3cm}
\caption{Pair production probabilities (black solid curves) in a bifrequent electric field with $\xi_1=1$, $\xi_2=0.1$, $N=7$ and $\omega=0.49072m$, as function of the relative phase between the field modes. 
The electron momenta in the top (bottom) panel are $p_x=0.25m$, $p_y=0$ ($p_x=0.01m$, $p_y=0.1m$).
The red dash-dotted lines show the corresponding Fourier coefficient $W_0$. The blue dotted (green dashed) curves show the results of the truncated Fourier sums in Eq.\,\eqref{sum} when terms up to $\ell =1$ (up to $\ell=2$) are taken into account.}
\label{figureSum}
\end{figure}

Figure~\ref{figureSum} shows the pair production probability $W(\vec p, \varphi)$ for fixed particle momenta as function of $\varphi$ (black solid curves). In general, the phase dependence can be rather involved, as exemplified in the top panel. In accordance with Eq.~\eqref{sum}, the probability can be decomposed into its Fourier components. The leading coefficient $W_0$ determines the phase-averaged value of probability (red dash-dotted line). By adding the next term with $\ell = 1$ in the Fourier expansion, the overall trend of $W(\vec p, \varphi)$ is reproduced roughly (blue dotted curve). When the $\ell=2$ term is taken into account as well, the approximate agreement becomes convincing (green dashed curve). Therefore, in what follows, we will concentrate on the first three terms in the Fourier series.

As illustrated in the bottom panel of Fig.~\ref{figureSum}, for certain particle momenta the picture simplifies considerably. In the example shown the shape of the pair production probability closely resembles a sin function and is pretty well approximated already by the first two Fourier components. The complex phase of the first Fourier coefficient $W_1$ in this case takes the value $\Phi_1\approx -\frac{\pi}{2}$,  accordingly.

\subsection{Relative Phase Contrast and Phase of the Phase}

In the framework of phase-of-the-phase spectroscopy, the relative-phase dependence of the pair production yield (as illustrated in Fig.~\ref{figureSum}) is encoded in a few functions of the particle momenta. As argued above, in the parameter regime under consideration, the absolute values of the Fourier coefficients $W_0(\vec p\,)$, $W_1(\vec p\,)$ and $W_2(\vec p\,)$ along with the complex phases $\Phi_1(\vec p\,)$ and $\Phi_2(\vec p\,)$ are sufficient to reconstruct the pair production signal with high accuracy. 

\begin{figure}[b]  
\vspace{-0.25cm}
\includegraphics[width=0.45\textwidth]{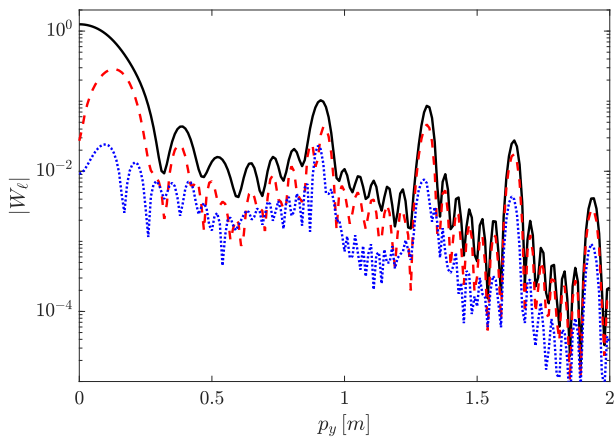}
\caption{Absolute values of the Fourier coefficients $W_0$ (black solid curve), $W_1$ (red dashed curve) and $W_2$ (blue dotted curve) associated with pair production in a bifrequent electric field with the parameters of Fig.~\ref{figureSum}. Shown is the dependence on the longitudinal electron momentum $p_y$, when the transverse momentum is zero.}
\vspace{-0.5cm} 
\label{figureRPC}
\end{figure}

The dependence of $|W_\ell(\vec p\,)|$ on the longitudinal electron momentum $p_y$, when the transverse momentum vanishes, is shown in Fig.~4 for $\ell=0$, 1 and 2. The field parameters are $\xi_1=1$, $\xi_2=0.1$, $N=7$ and $\omega=0.49072m$. The largest contribution results from $W_0$, which roughly follows the pair production probabilities shown in Fig.~\ref{figure1D}. Depending on the value of $\varphi$, the contributions from the terms with $\ell=1$ and $\ell=2$ in Eq.~\eqref{sum} either enhance or reduce the pair production yield at given $\vec p$. This leads to the differences between the curves in Fig.~\ref{figure1D}.

\begin{figure}[t]  
\begin{center}
\includegraphics[width=0.44\textwidth]{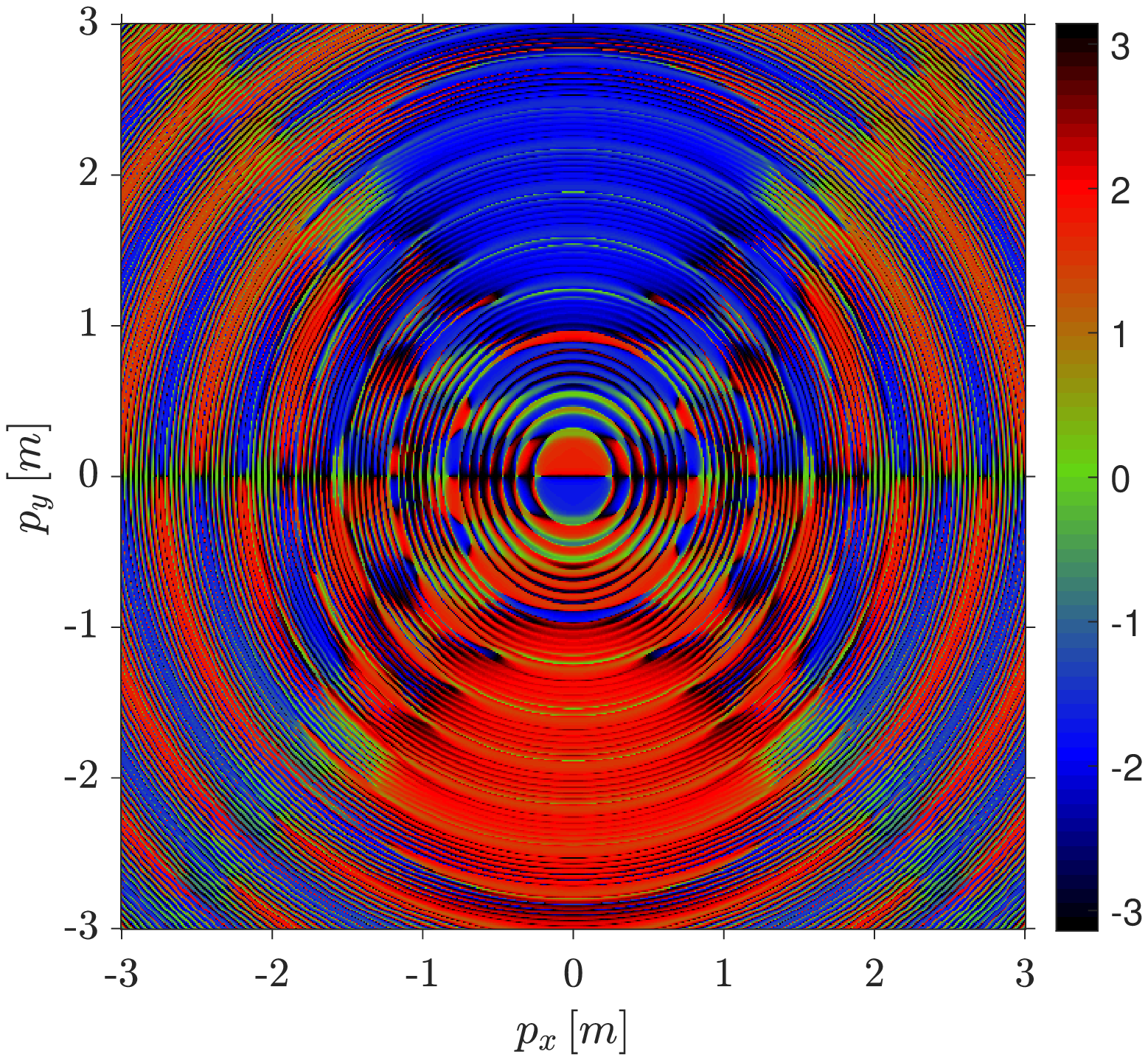}\\
\vspace{0.4cm} 
\includegraphics[width=0.44\textwidth]{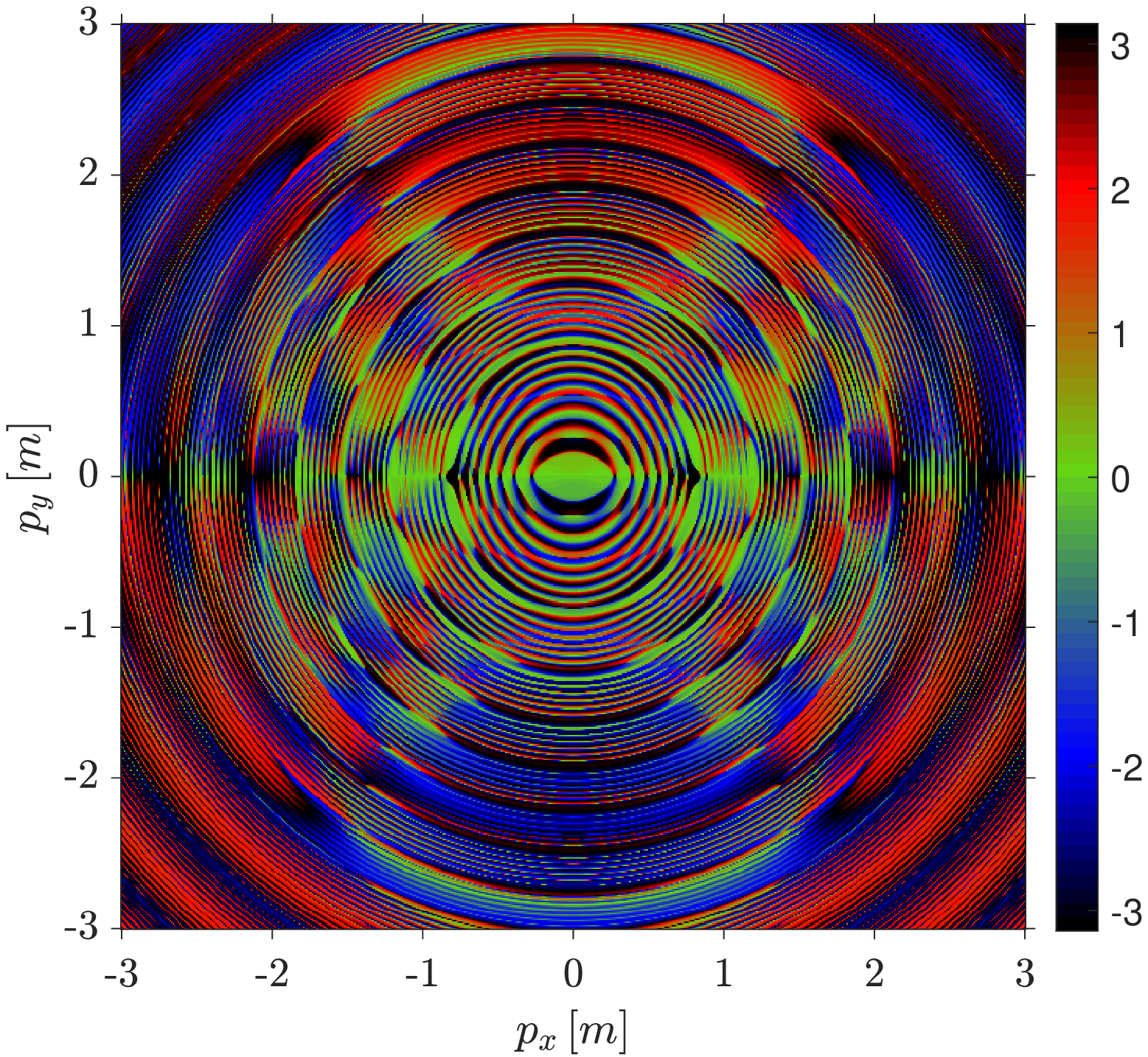}
\end{center}
\vspace{-0.5cm} 
\caption{Phase-of-the-phase spectra for the electron created in a bifrequent electric field with the parameters of Fig.~\ref{figureSum}. Top panel: $\Phi_1$, bottom panel: $\Phi_2$ (each measured in rad with $-\pi\le\Phi_\ell\le\pi$, as indicated by the color coding).}
\label{figurePoP}
\end{figure}

\begin{figure}[t]  
\begin{center}
\includegraphics[width=0.44\textwidth]{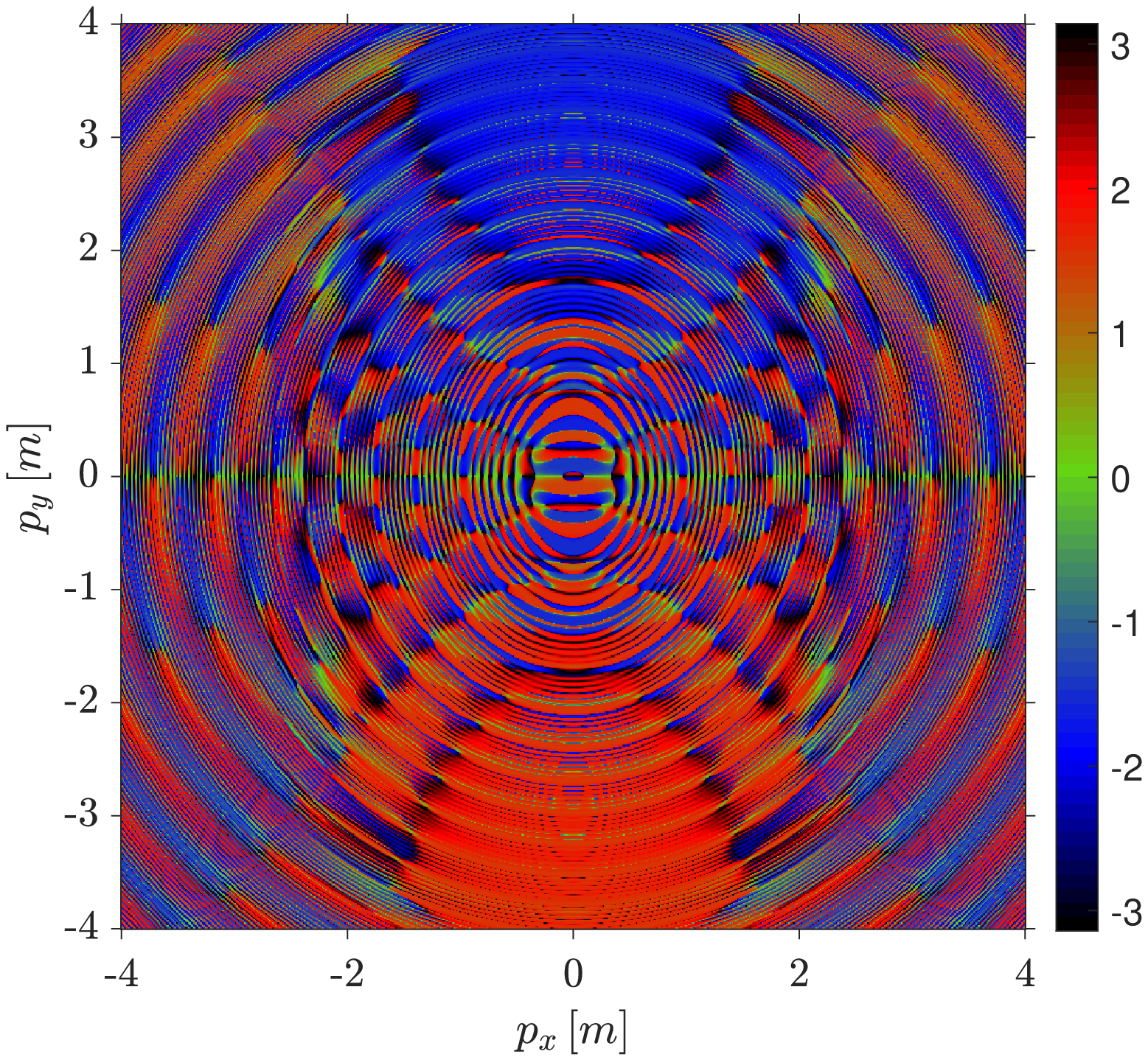}\\
\vspace{0.4cm} 
\includegraphics[width=0.44\textwidth]{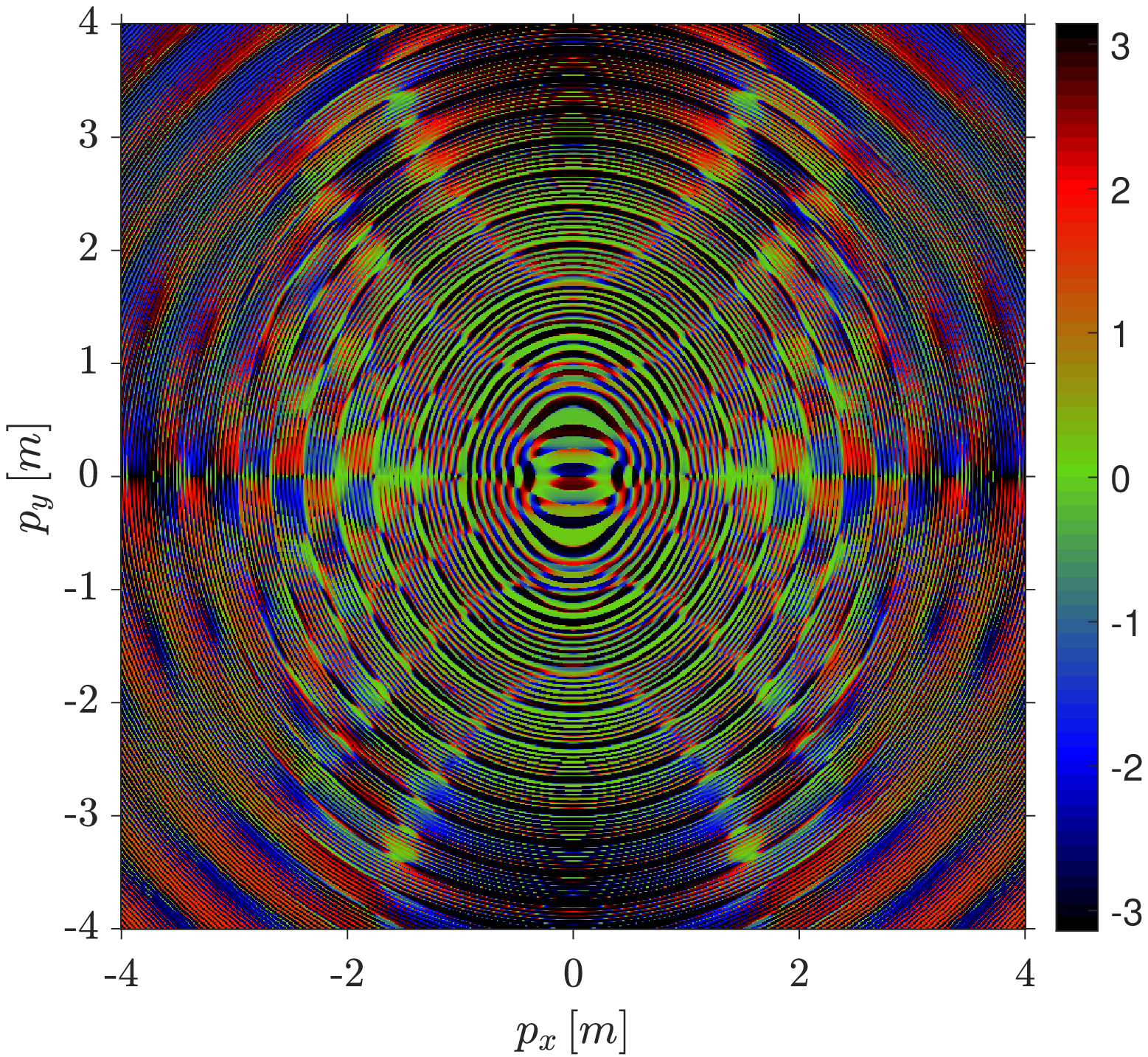}
\end{center}
\vspace{-0.5cm} 
\caption{Same as Fig.~\ref{figurePoP}, but with increased amplitude of the fundamental field mode, corresponding to $\xi_1=2$.}
\label{figurePoP2}
\end{figure}

Figures~\ref{figurePoP} and \ref{figurePoP2} show the phase-of-the-phase values $\Phi_1(\vec p\,)$ and $\Phi_2(\vec p\,)$ for the created electron in the $p_x$-$p_y$ plane for $\xi_1=1$ and $\xi_1=2$, respectively. A very complex structure is found, which is dominated by alternating red and blue areas. 
A characteristic checkerboard pattern arises this way, which closely resembles the structures found for ATI photoelectrons in bichromatic fields \cite{BauerPRL, BauerJPB, BauerPRA}. The spectra are symmetric under the transformation $p_x\to -p_x$ along the transverse direction, but asymmetric along the field direction. 
As a consequence, the corresponding spectra for the created positron would be inversed (i.e. $\Phi_\ell\to-\Phi_\ell$).

We first discuss the behavior of $\Phi_1$. The blue (red) areas belong to $\Phi_1\approx -\frac{\pi}{2}$ ($\Phi_1\approx +\frac{\pi}{2}$), corresponding to a $+$sin-like ($-$sin-like) dependence of the pair yield:
\begin{eqnarray}
W(\vec p, \varphi) &\approx& W_0(\vec p\,) + 2|W_1(\vec p\,)| \cos\left(\varphi\mp\frac{\pi}{2}\right)\nonumber\\
&=& W_0(\vec p\,) \pm 2|W_1(\vec p\,)| \sin(\varphi)\ .
\end{eqnarray}
Figures~\ref{figurePoP} and \ref{figurePoP2} show as general trend that, within a cone-shaped region around the field axis, a $+$sin-like dependence dominates for positive longitudinal momenta, $p_y>0$, and vice versa. This feature can be related to the shape of the underlying vector potential \eqref{A}, which is responsible for the pair production. In contrast to the case of $\varphi=0$, the vector potential is in general asymmetric when $\varphi\ne 0$. For example, when $\varphi$ lies between 0 and $\pi$, the maximum amplitude of $e{\vec A}(t)$ in positive $y$ direction exceeds its maximum amplitude in negative $y$ direction. The asymmetry of the vector potential leads to an asymmetry in the electron momentum spectra \cite{Brass, Alkofer3}.

This kind of relation was also found in Ref.~\cite{Krajewska} where the phase dependence of electron spectra resulting from pair production in the superposition of strong bichromatic laser and nuclear Coulomb fields were studied. It can be understood by noting that in strong-field processes the asymptotic longitudinal electron momentum (i.e. the momentum outside the field) is often related to $e{\vec A}(t_0)$ where $t_0$ denotes the ``moment'' when the electron has entered into the field. In line with this picture we see that, in the present situation, positive $p_y$ values are favored, whereas the production probability for electrons with negative $p_y$ component has a tendency to be reduced when $\varphi$ grows from 0 towards positive values. The same general trend was found in phase-of-the-phase spectra of ATI photoelectrons \cite{BauerJPB}. In our case the transfer of asymmetry from the vector potential (whose magnitude is limited by $|e{\vec A}(t)|\lesssim m$ for the chosen parameters $\xi_2\ll\xi_1\sim 1$) to the electron momenta is mediated by the quasi-energy \eqref{epsilon}, which contains the combination $p_y-eA(t)$. Thus, when the maximum of $eA(t)$ in positive direction exceeds the maximum in negative direction, it is ``easier'' to produce electrons with rather large positive than with rather large negative $p_y$ values. For positrons the situation is reversed: here, a $+$sin-like dependence dominates for negative longitudinal momenta.\\

Outside of the cone-shaped region, blue and red areas alternate frequently along the resonance rings, leading to a pronounced checkerboard pattern. Accordingly, when moving along (or across) a resonance ring, a sign change from $+\sin$-like to $-\sin$-like (or vice versa) occurs. As argued in \cite{BauerJPB}, the appearance of such a pattern is related to a redistribution of probability in the $p_x$-$p_y$ plane, when the relative phase changes. This means, the increase of probability in some regions is accompanied by a decrease of probability in neighboring regions. In the present case of pair production, this redistribution can be caused by the $\varphi$-dependence of the quasi-energy \eqref{epsilon}. For the chosen parameters, the latter changes by $\sim 10^{-2}m$ when $\varphi$ varies from 0 to $2\pi$. While being small, this change lies in the same order of magnitude as the widths of the multi\-photon resonances \cite{Mocken}. Consequently, when $\varphi$ is varied, one may need either larger or smaller momentum values to approach the resonance condition.

The lower panels in Figs.~\ref{figurePoP} and \ref{figurePoP2} illustrate the behavior of $\Phi_2$. The overall appearance resembles the phase-of-the-phase $\Phi_1$, but the structure has become even more rich. In addition to the cone-shaped regions and the blue-and-red checkerboard pattern, there are now also many green areas which correspond to $\Phi_2\approx 0$ and, thus, a $+\cos$-like dependence with $\cos(2\varphi)$. Besides, one may notice that the colors are exchanged within the cone-shaped regions: now a $+\sin$-like behavior dominates for $p_y<0$. Hence, the second term in the Fourier series ($\ell = 2$) seems to partially counteract the influence of the first term ($\ell = 1$). In general, however, the contribution of the second term is substantially smaller, as was shown in Fig.~\ref{figureRPC}.

\section{Conclusion and Outlook}

Electron-positron pair production from vacuum in strong bifrequent electric fields was studied. The influence of the relative phase between a fundamental field mode and its second harmonic was analyzed by phase-of-the-phase spectroscopy, decomposing the pair yield into its corresponding Fourier components. We have shown that the phase-of-the-phase spectra for the created electron closely resemble the corresponding outcomes from strong-field photoionization which have been obtained previously \cite{BauerJPB}. The spectra for the created positron differ by an overall sign. For the applied field parameters the pair production signal can be well reconstructed by inclusion of the first three Fourier terms.

Phase-of-the-phase spectroscopy has been introduced in strong-field atomic physics as a means to experimentally discriminate photoelectrons emitted via the coherent interaction with a two-color laser field from those electrons which result from incoherent and, thus, phase-independent processes (such as thermal emission or collisional ionization via incoherent scatterings) \cite{BauerPRL, BauerJPB, BauerPRA}. 
When applied to strong-field pair production the method could similarly help to distinguish the desired signal of coherently produced pairs from the -- potentially strong -- background noise which might result from other processes, such as collisions between residual atoms and laser-accelerated electrons in the rest gas, for instance. One should mention, though, that such an application -- at the very high field intensities required for pair production -- certainly represents a major technical challenge. Most likely, it would therefore become relevant only after pair-production experiments at the upcoming high-field facilities \cite{ELI, XCELS, XFEL} have become a routine. 

The method is not limited to the scenario of the present paper. It can also be applied to pair production processes in other field configurations, such as high-intensity laser beams combined with $\gamma$-ray photons \cite{Narozhny, Jansen} or nuclear fields \cite{DiPiazza, Krajewska, Augustin, Roshchupkin}. Besides, it is applicable not only to the relative phase of a bichromatic field, but to any continuous variable which the field depends periodically on and which can be controlled in experiment (such as the carrier-envelope phase of a few-cycle laser pulse \cite{Paulus,Zherebtsov}).

\section*{Acknowledgement}

This work has been funded by the Deutsche Forschungsgemeinschaft (DFG) under Grant No. 416699545 within the Research Unit FOR 2783/1.



\begin{thebibliography}{33}
\bibitem{dipole} In the realm of nonrelativistic laser-atom interactions, the dipole approximation to the laser fields, which neglects their spatial dependencies, is usually well justified.
\bibitem{q_0} The quasi-energy of the created positron, which formally follows from Eq.~\eqref{epsilon} by the replacements $\vec p\to-\vec p$ and $e\to-e$, attains the same value.
\end{thebibliography}
\end{document}